\documentclass[preprint,showpacs,preprintnumbers,amsmath,amssymb,nofootinbib]{revtex4-1}

\usepackage{graphicx}
\usepackage{dcolumn}
\usepackage{bm}

\newcommand{\beq}{\begin{equation}}
\newcommand{\eeq}{\end{equation}}
\newcommand{\beqy}{\begin{eqnarray}}
\newcommand{\eeqy}{\end{eqnarray}}
\newcommand{\beqyn}{\begin{eqnarray*}}
\newcommand{\eeqyn}{\end{eqnarray*}}
\newcommand{\nl}{\newline}

\newcommand{\bc}{\begin{center}}
\newcommand{\ec}{\end{center}}
\newcommand{\bmin}{\begin{minipage}}
\newcommand{\emin}{\end{minipage}}

\begin{document}

\title{Reply to the comment of Huey-Wen Lin and Keh-Fei Liu on ``Controversy concerning the definition of quark and gluon
    angular momentum'' by E. Leader (PRD 83, 096012 (2011)) }

\author{Elliot Leader}
 \email{e.leader@imperial.ac.uk}
\affiliation{Blackett laboratory \\Imperial College London \\ Prince Consort Road\\ London SW7 2AZ, UK}

\date{\today}


 \begin{abstract}Lin and Liu evaluate the nucleon expectation value of the non gauge-invariant canonical quark momentum operator on a lattice, and obtain  zero. They conclude that my argument that, despite the non gauge-invariance of the operator, its physical matrix elements \emph{are} gauge-invariant,  cannot be correct. I show that their result has no bearing on the question of gauge-invariance, and also point to an amusing lattice paradox.
\end{abstract}
\maketitle

In the controversy concerning the definition of quark and gluon momentum operators \cite{Leader:2011za} I suggested that it is the generator of translations which should be identified as the linear momentum operator, and this implies using the \emph{canonical} version  $\bm{P}_{\textrm{can}}$ as the momentum operator. This operator itself is not gauge-invariant, but what is important are its matrix elements between physical states, and I claimed to show that these are, in fact, gauge-invariant. To test this claim Lin and Liu (LL) carried out a connected insertion (CI) lattice calculation of the nucleon expectation value of the $Z$-component of  $ \langle \bm{P}_{\textrm{can}}(\textrm{quark}) \rangle $ and of the corresponding gauge-invariant Bellinfante version $\langle \bm{P}_{\textrm{bel}}(\textrm{quark}) \rangle $ in QCD, arguing that if an expectation value is non gauge-invariant then its lattice value will be zero. They obtained

\beq \label{res} \langle \bm{P}_{\textrm{can}}(\textrm{quark})\rangle_{\textrm{LatCI}} \approx 0 \qquad \langle \bm{P}_{\textrm{bel}}(\textrm{quark}) \rangle_{\textrm{LatCI}} \approx 0.42  \eeq

from which they conclude that $\langle \bm{P}_{\textrm{can}}(\textrm{quark}) \rangle $  cannot be gauge-invariant. \nl
I shall argue that, on the one hand, their result raises an interesting question about lattice methods, and, on the other, that the result cannot be used to deduce the non gauge-invariance of the canonical momentum expectation value. \nl
Now it is known that the disconnected insertions (DI) in the lattice calculations provide a small fraction  of the quark momentum, so it is unlikely that the LL conclusion will change after inclusion of the DI. Therefore,  we may ignore this possibility, and explaining the LL result then raises several intriguing issues. \nl
 \emph{A lattice paradox}:  Since also the canonical \emph{gluon} momentum operator is not gauge-invariant we must suppose that the lattice value
$ \langle \bm{P}_{\textrm{can}}(\textrm{gluon}) \rangle_{\textrm{LatCI}} $ will turn out to be zero. Thus, for the total momentum

\beq \label{tot} \langle \bm{P}_{\textrm{can}}(\textrm{total}) \rangle_{\textrm{LatCI}} =\langle \bm{P}_{\textrm{can}}(\textrm{quark}) \rangle_{\textrm{LatCI}} + \langle \bm{P}_{\textrm{can}}(\textrm{gluon}) \rangle_{\textrm{LatCI}} \eeq
one will obtain zero. However $\langle \bm{P}_{\textrm{can}}(\textrm{total}) \rangle_{\textrm{Lat}} $ cannot be zero since it is known that

\beq \label{canbel} \langle \bm{P}_{\textrm{can}}(\textrm{total}) \rangle = \langle \bm{P}_{\textrm{bel}}(\textrm{total}) \rangle  \eeq
because the \emph{total} momentum operators differ by the integral of a divergence. \nl
Thus we are faced with the peculiar conclusion that

\beq \label{latsick} \sum \left \{ \, (\textrm{matrix elements})_{\textrm{Lat}}\right \} \neq \left \{ \sum (\textrm{matrix elements})\right \}_{\textrm{Lat}}. \eeq

\emph{Why} $\langle \bm{P}_{\textrm{can}}(\textrm{quark})\rangle_{\textrm{Lat}}= 0 \nRightarrow $ \emph{non gauge-invariance}: The canonical momentum density  given by

\beq \label{Tcan}   t^{0j}_{can} = \frac{i}{2}\, \bar{\psi}\gamma^0 \overleftrightarrow{\partial}^j \,\psi  \eeq
is a strictly local operator. However, to implement the derivative on a lattice it is necessary to utilize finite difference techniques, so that $ t^{0j}_{can}$ on the lattice involves products of fermion fields at different space-time points. What is being evaluated by LL, therefore, is closely analogous to the Path Integral evaluation of the vacuum expectation value $ \langle \, 0 \, | \, \psi (x) \bar{\psi}(y)\, | \, 0 \, \rangle_{\textrm{PathInt}} $, which, as demonstrated elegantly in Section 2.12 of \cite{Collins:85}, vanishes when integrating over gauge configurations in an unconstrained way. Thus the vanishing of $\langle \bm{P}_{\textrm{can}}(\textrm{quark})\rangle_{\textrm{Lat}}$ is entirely a consequence of the lattice implementation of the derivative, and tells us nothing about the issue of gauge-invariance.

\bibliography{Elliot_General}

\end{document}